\documentstyle[12pt]{article}
\newlength{\inda}
\newlength{\indb}
\def\Re{I\!\!\!\,R}

\begin{document}

\markright{Cylindrical waves and spinning strings}

\begin{center}
{\large \bf Asymptotic behaviour of cylindrical waves
interacting with spinning strings}
\\[3mm] {Nenad Manojlovi\'c}
\\[3mm] {\it Universidade do Algarve, Campus de Gambelas, 8000 Faro, Portugal}
\\[3mm] {Guillermo A Mena Marug\'{a}n}
\\[3mm] {\it  Instituto de Matem\'{a}ticas y F\'{\i}sica Fundamental, CSIC,
\\ Serrano 121, 28006 Madrid, Spain}
\\[3mm] 24 March 2001
\end{center}

\begin{abstract}

We consider a family of cylindrical spacetimes endowed with
angular momentum that are solutions to the vacuum Einstein
equations outside the symmetry axis. This family was recently
obtained by performing a complete gauge fixing adapted to
cylindrical symmetry. In the present work, we find boundary
conditions that ensure that the metric arising from this gauge
fixing is well defined and that the resulting reduced system has
a consistent Hamiltonian dynamics. These boundary conditions
must be imposed both on the symmetry axis and in the region far
from the axis at spacelike infinity. Employing such conditions,
we determine the asymptotic behaviour of the metric close to and
far from the axis. In each of these regions, the approximate
metric describes a conical geometry with a time dislocation. In
particular, around the symmetry axis the effect of the
singularity consists in inducing a constant deficit angle and a
timelike helical structure. Based on these results and on the
fact that the degrees of freedom in our family of metrics
coincide with those of cylindrical vacuum gravity, we argue that
the analysed set of spacetimes represent cylindrical
gravitational waves surrounding a spinning cosmic string. For
any of these spacetimes, a prediction of our analysis is that
the wave content increases the deficit angle at spatial infinity
with respect to that detected around the axis.

\end{abstract}

\renewcommand{\theequation}{\arabic{section}.\arabic{equation}}

\section{Introduction}

The study of cylindrically symmetric spacetimes has received a
lot of attention in general relativity \cite{kra,ver}.
Cylindrical gravity possesses a non-trivial field content, and
therefore provides a suitable arena where one can test and
develop methods (such as solutions generating techniques
\cite{ver} and quantisation procedures \cite{ku,api}) which are
capable of dealing with the infinite number of degrees of
freedom of the gravitational theory. More importantly,
cylindrical spacetimes have found application in describing
(idealized) situations of interest in gravitational physics and
astrophysics. One of these applications is the analysis of the
propagation and interaction of gravitational waves.

The first family of exact solutions corresponding to
time-dependent cylindrical waves in a vacuum seems to have been
obtained by Beck \cite{be}. This family was rediscovered by
Einstein and Rosen in a systematic analysis of spacetimes which
represent cylindrical or plane gravitational waves \cite{ero}.
The waves considered by Einstein and Rosen display what has been
called `whole-cylinder symmetry' \cite{mel,tho}, namely, they
are not only cylindrically symmetric, but also linearly
polarized, so that the metric can be written globally in a
diagonal form \cite{kra}. The most general vacuum solution
describing cylindrical waves was studied independently by
Kompaneets and by Ehlers {\it et al.} \cite{ko}. These waves can
be described by a gravitational model whose configuration space
has two field-like degrees of freedom which are subject to
generalized wave equations \cite{roto}. The general solution to
such dynamical equations is not explicitly known.

A different type of physical phenomena that can be associated
with cylindrical spacetimes are cosmic strings \cite{vi}. Cosmic
strings are topological defects (characterized by a non-trivial
homotopy group $\pi_1$) that could be formed during phase
transitions predicted by grand unified theories \cite{kib}. In
the particular case of a straight cosmic string with a
non-vanishing mass per unit length, the exterior gravitational
field is simply a vacuum cylindrical spacetime that presents a
conical defect \cite{vi,go}. The string may also possess spin,
providing the exterior spacetime with an angular momentum
\cite{djt,maz}. The existence of straight cosmic strings would
have important consequences for astrophysics: they would give
rise to discontinuities in the microwave background \cite{step}
and act as gravitational lenses \cite{lense}. It has also been
argued that cosmic strings might have produced the density
fluctuations that led to galaxy formation \cite{seed}.
Nevertheless, recent observations indicate that the anisotropy
of the cosmic microwave background has most probably originated
in an inflationary scenario \cite{boom}, although the
possibility that cosmic strings are partially responsible for
the formation of structure is still open \cite{def}.

In contradistinction to the properties displayed by cylindrical
waves, whose spacetime is regular everywhere, a straight cosmic
string can be described by a cylindrical spacetime whose
symmetry axis is singular (quasi-regular, to be more precise
\cite{qua}). This singularity corresponds to the line source
provided by the string \cite{sou,gal}. The exterior is a conical
geometry that possesses a deficit angle and an angular momentum
which are proportional, respectively, to the mass and spin of
the string \cite{djt,maz}.

Another example of cylindrical spacetimes are those that
represent cylindrical waves surrounding a straight cosmic
string. The analysis of these spacetimes allows a simplified
discussion of the interaction between gravitational waves and
strings. A few (parameter-dependent) families of solutions of
this type have been found explicitly \cite{ver,xan,etso,vega}.
Essentially, these solutions have been constructed by applying
the soliton technique of Belinski\v{\i} and Zakharov, which
makes use of the existence of two commuting Killing vectors
\cite{ver,beza}. The spacetimes obtained in this way generally
present, at spatial infinity and/or around the line source, a
dislocation in the direction of the axis \cite{gal,dislo}.
Exceptions are the Garriga-Verdaguer solutions \cite{vega},
whose metric is diagonal, and a subfamily of the spacetimes
discussed in \cite{etso}. A property that is shared by all of
these solutions, as well as by the gravitational cylindrical
waves and the straight cosmic string without spin (but not by
the spinning string), is that the surface spanned by the
rotational and translational Killing vectors admits an
orthogonal surface, i.e. that the considered isometry group is
orthogonally transitive \cite{kra}. In practice, this implies
the vanishing of the metric components that mix the radial
coordinate or the timelike one with the coordinates that
describe the Killing trajectories.

One of the authors has recently considered all possible
cylindrically symmetric spacetimes that are solutions to the
vacuum Einstein equations outside the axis, allowing them to
contain a non-zero angular momentum \cite{me}. Note that
cylindrical waves and straight cosmic strings are included in
this family of solutions. For all of the spacetimes in this
family, it has actually been shown that, once the radial
coordinate is chosen orthogonal to the group trajectories, the
shift vector has non-vanishing projection on the Killing orbits
provided that the angular momentum differs from zero. Therefore,
orthogonal transitivity does not hold in this case. In addition,
it has also been shown that, in any of these solutions, there
exists a dynamically conserved quantity that describes the total
energy per unit length in the axis direction \cite{me}. A
similar energy density was known to exist for purely
gravitational cylindrical waves \cite{roto,ama,va}. The energy
density for those waves is actually a non-polynomial function of
the total C-energy introduced by Thorne \cite{tho}, and turns
out to be positive and bounded from above. Although the total
C-energy ceases to be a constant of motion when the angular
momentum differs from zero, the results of \cite{me} generalize
the expression of the energy in the presence of an axial
singularity endowed with spin. In particular, the energy density
continues to be bounded both from below and above.

In this paper, we will argue that, apart from a possible
dislocation, the spacetimes considered in \cite{me} can, in
fact, be interpreted as describing the interaction of a spinning
cosmic string with a cylindrical gravitational wave. In this
sense, it is worth noting that the degrees of freedom outside
the symmetry axis are just those corresponding to the
cylindrical reduction of vacuum general relativity, namely, the
field-like degrees of freedom of cylindrical waves. Furthermore,
we will prove that the singular behaviour allowed on the axis
produces a conical geometry which is endowed with a constant
angular momentum and whose deficit angle does not vary in time.
These are precisely the effects of a spinning cosmic string. So
the intuitive picture that one gets is that of a stringy defect
surrounded by an empty cylindrical spacetime which differs from
the Minkowskian vacuum in that it generally contains
time-dependent gravitational fields.

In order to attain this picture, we will investigate the form of
the metric, both at spatial infinity and around the symmetry
axis. We will show that these asymptotic metrics can indeed be
understood as being created by a spinning cosmic string
interacting in a non-linear way with a cylindrical wave. We will
also see that the main effect of the gravitational content at
spatial infinity is to increase the deficit angle caused by the
string. To reach these conclusions, we will first carefully
determine the boundary conditions which guarantee that the
analysed cylindrical solutions are rigorously defined. These
boundary conditions are crucial to show that the system
possesses non-trivial solutions that are physically acceptable
and admit a well defined notion of energy density. If such
boundary conditions did not exist, our whole analysis would
simply remain as a formal discussion devoid of physical content.
Furthermore, it is precisely the knowledge of these boundary
conditions what will allow us to obtain the asymptotic behaviour
of the metric and discuss the physical phenomena that can take
place in spacetime.

The paper is organized as follows. In section 2, we summarize
the main results of \cite{me}. This includes the expression of
the metric in terms of the field-like degrees of freedom of the
system, the dynamical equations that these fields satisfy and
the Hamiltonian that generates the evolution. In section 3 we
discuss boundary conditions that guarantee that the metric
expressions are well defined. The stability of these conditions
is studied in section 4. Once a set of dynamically stable
conditions has been determined, we check in section 5 that they
really lead to a consistent Hamiltonian formalism. We then
analyse the behaviour of the metric at large and short distances
from the axis in sections 6 and 7, respectively. In section 7,
we also discuss the physical interpretation of the solutions.
Section 8 contains the conclusions and further discussion.
Finally, two appendices are added.

\section{Spinning spacetimes}

In this section, we briefly review the main results of \cite{me}
concerning the Hamiltonian formalism for cylindrical spacetimes
in vacuo that, in principle, do not include the axis of
symmetry. The starting point of this analysis is the Hamiltonian
formulation of vacuum general relativity for spacetimes with two
commuting Killing vectors, one of them translational and the
other one rotational. It is possible to introduce coordinates
$x^a=\{z,\theta\}$ (with $z\in\Re$, $\theta\in S^1$ and $a=1,2$)
adapted to these Killing isometries so that the metric is
independent of $x^a$. In addition, we will call $x^3=r>0$ the
radial coordinate, denote spatial indices with Latin letters
from the middle of the alphabet and adopt units such that
$4G=c=1$, where $G$ is the Newton constant per unit length in
the $z$ direction.

All of the gravitational constraints can be removed from the
system by imposing a gauge fixing that makes use of the
cylindrical symmetry. Let us first consider the momentum
constraints associated with the Killing vectors, which have the
form ${\cal H}_a=-2\partial_r(h_{ai}\Pi^{ir})$. Here, $h_{ij}$
is the induced 3-metric and $\Pi^{ij}$ its canonically conjugate
momentum \cite{me}. These constraints can be eliminated by
requiring that the metric components $h_{ar}$ vanish. In this
gauge, the solution to the constraints ${\cal H}_a=0$ is given
by $\Pi^{ar}=h^{ab}c_b/4$ ($a,b=1,2$), where the $c_a$s are two
constants related to global properties of the spacetime. In
particular, if the spacetime is regular everywhere, including
the axis, the constants $c_a$ must vanish. The stability of the
gauge conditions, on the other hand, determines the value of the
components $N^a$ of the shift vector. After this partial gauge
fixing, the system has two constraints and the gravitational
degrees of freedom reside in the components $h_{ab}$ and
$h_{rr}$ of the induced metric.

In fact, the constants $c_z$ and $c_{\theta}$ describe,
respectively, the linear momentum in the $z$ direction and the
angular momentum contained in the spacetime, both of them
expressed as linear densities. In order to prove this statement,
it suffices to remember that, modulo surface terms, the
generators of the asymptotic rotations (far from the axis) and
the asymptotic translations in the $z$ direction both have the
integral form $\int dr N^i {\cal H}_i$, where the shift $N^i$
vanishes in the region $r\ll 1$ and tends to $\delta_a^i$ when
$r\rightarrow\infty$, with $a=1$ for translations and $a=2$ for
rotations \cite{ama}. Taking into account the expression of
${\cal H}_a$, a simple integration by parts shows that the
considered generators are differentiable on the gravitational
phase space, and hence well defined, provided that the neglected
surface terms coincide with the limit of $2h_{ai}\Pi^{ir}$ when
$r$ tends to infinity. With our gauge fixing, this limit is half
the constant $c_a$. Therefore, on solutions to the gravitational
constraints, the values of the analysed generators, which are
the linear and angular momentum densities under consideration
\cite{ama,he}, are given in our system of units by the constant
quantities $c_z/2$ and $c_{\theta}/2$, respectively. In the
following, we will allow the possibility that the cylindrical
spacetime contains a non-zero angular momentum, but will
restrict our attention to the case in which $c_z$ vanishes.
Otherwise, our gauge-fixing procedure would lead to metric
expressions plagued with divergences that would render them
meaningless \cite{me}.

The gauge freedom corresponding to the only momentum constraint
that remains on the system can be removed by choosing as a
radial coordinate the square root of the determinant of the
metric on Killing orbits, since this metric degenerates on the
axis and its determinant is supposed to possess a spacelike
gradient. The canonical momentum of this determinant can then be
determined by solving the radial constraint on the gauge
section. The requirement of stability of the gauge condition, on
the other hand, fixes the radial component of the shift, which
turns out to be proportional to the momentum $P_w$ canonically
conjugate to the metric variable $w=\ln{\sqrt{h_{zz}h_{rr}}}$.
One can then make the shift component $N^r$ equal to zero by
imposing that $P_w$ vanishes as a gauge condition for the
Hamiltonian constraint. In this way, one eliminates all the
gauge degrees of freedom from the system. The expression of the
metric function $w$ can be obtained by solving the Hamiltonian
constraint with $P_w=0$. In addition, the dynamical stability of
the gauge (i.e. $\dot{P}_w=0$, where the overdot denotes the
time derivative) determines the value of the lapse function,
assuming that it approaches the unity in the limit
$r\rightarrow\infty$.

After this complete gauge fixing, the system can be described by
two canonically conjugate pairs of field-like degrees of
freedom, which we will call $(v,P_v)$ and $(y,P_y)$. The reduced
metric can be written in the form
\begin{equation}\label{met}
ds^2=e^{2w+y}\left[-\bar{N}^2dt^2+dr^2\right]+
e^{y}r^2(d\theta+N^{\theta}dt)^2+e^{-y}
\left[dz-vd\theta+(N^z\!-\!vN^{\theta})dt\right]^2,
\end{equation}
where
\begin{eqnarray}\label{w}
&&e^{2w}=\frac{4\bar{E}[r]r^2}{c_{\theta}^2+4Dr^2-2
c_{\theta}^2r^2\int_0^r ds\,s^{-3}\, (\bar{E}[s]-1)},\\
\label{NFE} &&\bar{N}=F_{\infty}e^{-2w}\bar{E}[r],\hspace*{.5cm}
F_{\infty}=\frac{e^{w_{\infty}}}{\bar{E}_{\infty}},\hspace*{.5cm}
\bar{E}[r]=\exp{\left(\int_0^r\bar{H}\right)},\\
 \label{barH}
&&\bar{H}=\frac{2}{r}\left[\frac{(r \partial_ry)^2}{4}\!+\!
\frac{(\partial_rv)^2}{4}e^{-2y}+P_y^2+P_v^2r^2e^{2y}\right],
\end{eqnarray}
and the shift vector is
\begin{equation}\label{shift}
N^{\theta}=c_{\theta}F_{\infty}\left\{\frac{1}{2r^2}+
\int^{\infty}_r\frac{ds}{s^3}(\bar{E}[s]-1)
\right\},\hspace*{.5cm}
N^{z}=c_{\theta}F_{\infty}\int^{\infty}_{r}\frac{ds}
{s^3}\,v\,\bar{E}[s].\end{equation}

In these equations, $w_{\infty}$ and $\bar{E}_{\infty}$ are,
respectively, the values taken by $w$ and $\bar{E}[r]$ in the
limit $r\rightarrow\infty$. To arrive at these expressions, the
shift vector has been chosen to vanish when $r$ tends to
infinity and the value of $y$ in this limit has been set equal
to zero by an appropriate scaling of $r$, $z$, $c_{\theta}$ and
the fields $(v,P_v)$. Finally, the parameter $D$, which
determines the sub-leading terms in the metric function $e^{2w}$
around the axis, has been assumed to be a positive constant.
This assumption is necessary if one wants that the dynamics of
the gauge-fixed system is generated (at least formally) by a
Hamiltonian \cite{me}. We will show in section 5 that the
constancy of $D>0$ is, in fact, guaranteed by the boundary
conditions that the fields must satisfy.

The dynamical evolution is dictated by the equations
\begin{eqnarray}\label{vdot}
&&\dot{v}=2F_{\infty}P_v r e^{2y-2w}\bar{E}[r],\\ \label{ydot}
&&\dot{y}=2F_{\infty}\frac{P_y}{r}e^{-2w}\bar{E}[r],\\
\label{Pvdot}
&&\dot{P}_v= F_{\infty}\partial_r\!\left(\frac{ \partial_rv}{2r}
e^{-2y-2w}\bar{E}[r]\right),\\ \label{Pydot}
&&\dot{P}_y=F_{\infty}\partial_r\!\left(\frac{\partial_ry}{2}r
e^{-2w}\bar{E}[r]\right)-\frac{F_{\infty}}{2r}e^{-2w}
\bar{E}[r]\left[4P_v^2r^2e^{2y}- ( \partial_rv)^2e^{-2y}\right],
\end{eqnarray}
which are generated in the reduced system, via Poisson brackets,
by the Hamiltonian $H_R=1-e^{-w_{\infty}}$. Employing equation
(\ref{w}), one explicitly obtains
\begin{equation}\label{Hom}
H_R=1-\sqrt{\frac{2D-c_{\theta}^2\,\Omega}{2\bar{E}_{\infty}}},
\hspace*{.5cm}\Omega=\int^{\infty}_0\frac{dr}{r^3}
\left(\bar{E}[r]-1\right).\end{equation}
We are assuming that the integrals involved in these expressions
converge; we will return to this point in section 3. An additive
constant in the Hamiltonian has been fixed by requiring that the
energy vanishes for Minkowski spacetime (i.e. when
$D=\bar{E}_{\infty}=1$, $c_{\theta}=0$).

Note that, since the reduced Hamiltonian must be real to define
an acceptable time evolution, the functional expression
(\ref{Hom}) implies that the value of $H_R$, which provides the
energy density per unit length in the axis direction, is bounded
from above by unity. The value $H_R=1$ cannot be reached,
because when $e^{-w_{\infty}}$ vanishes the metric is
ill-defined, according to our equations. On the other hand,
taking into account that $\bar{H}$ is a non-negative function on
the reduced phase space, one can see that $\bar{E}[r]\geq 1$ for
all values of $r>0$. As a particular consequence, $\Omega$ is
non-negative. It is then easy to find a lower bound for the
energy density, provided that the positive parameter $D$ is
fixed. One gets that $H_R\geq (1-\sqrt{D})$. The lower bound is
reached on solutions with vanishing momenta $P_v$ and $P_y$ and
constant fields $v$ and $y$, because it is only then that
$\bar{H}$ vanishes and $\bar{E}[r]=1$. Remembering that the
value of $y$ when $r\rightarrow\infty$ has been set equal to
zero and assuming that there exists no dislocation in the
direction of the axis \cite{gal,dislo}, we are then left with
only one spacetime. This spacetime, which minimizes the energy
density and can therefore be regarded as a background for the
solutions with parameters $D$ and $c_{\theta}$, is precisely the
region without closed timelike curves (CTCs) in the exterior of
a spinning cosmic string \cite{me}.

As we have briefly commented above, all the integrals over
$r\in\Re^+$ involved in the metric expressions must converge;
otherwise, our previous calculations would lead to physically
unacceptable solutions. Supposing that the fields
$\{v,y,P_v,P_y\}$ are sufficiently smooth as functions of $r$
over the positive axis, the convergence can be ensured by
imposing suitable boundary conditions on the fields, both at
$r=0$ and at infinity. We will discuss this point in section 3.
Of course, such boundary conditions must hold at all times and,
therefore, be preserved by the dynamical evolution. This issue
will be analysed in section 4. In addition, it is clear from the
form of the metric that, for real fields, the spacetime is
Lorentzian with $t$ being the time coordinate if and only if
$e^{2w}$ is positive for all values of $r$. In fact, remembering
that $\bar{H}$ is a non-negative function on phase space and
that $D>0$, one can check from equation (\ref{w}) that $e^{2w}$
is strictly positive in $r>0$ provided that $e^{2w_{\infty}}$ is
positive \cite{me}. This last requirement amounts to the
condition $2D>c_{\theta}^2\Omega$. The only effect of this
inequality when $c_{\theta}\neq 0$ is to restrict the admissible
initial data for the fields $\{v,y,P_v,P_y\}$. The reason is
that $e^{2w_{\infty}}$ is a constant of motion, because it does
not depend explicitly on time ($D$ is a constant) and commutes
with the reduced Hamiltonian $H_R=1-e^{-w_{\infty}}$ under
Poisson brackets \cite{me}. So $e^{2w_{\infty}}$ remains
positive in the evolution if it is originally positive. Hence,
the inequality $2D>c_{\theta}^2\Omega$ needs only be imposed at
a given, initial time. Note also that $\Omega$ vanishes if so do
all the fields. Consequently, for every value of $c_{\theta}$
and $D>0$, there exists a region on phase space around the
origin $v=y=P_v=P_y=0$ where the considered inequality is
satisfied.

On the other hand, accepting that the dynamical equations for
our fields are valid not only in the region $r>0$, but also when
one approaches the axis, a consistency condition that must be
verified is the constancy of the parameter $D$. Finally, if the
considered spacetimes admit a reduced Hamiltonian formulation,
it is necessary that the Hamiltonian be not only finite for
physical solutions, but also differentiable on the corresponding
phase space. All of these extra requirements will be analysed in
section 5, where we will check that they are satisfied as a
result of the boundary conditions imposed on the fields.

So, in summary, the situation is as follows. Although we have
obtained a formal expression for the gauge-fixed metric and
determined the equations of motion that satisfy the degrees of
freedom, the existence of solutions with a rigorously defined
metric is possible only if we can find suitable boundary
conditions compatible with the dynamics and the constancy of
$D$. Moreover, these solutions correspond to physically
acceptable spacetimes only if the restriction
$2D>c_{\theta}^2\Omega$ is satisfied by the initial data.
Finally, the reduced model possesses a well defined Hamiltonian
formalism and a constant of motion that provides a valid notion
of energy density only if the reduced Hamiltonian is
differentiable. Hence, the importance of proving that the system
admits a set of satisfactory boundary conditions and analysing
their consequences.

\section{Convergence of the integrals}
\setcounter{equation}{0}

In the rest of the paper, we will assume that the fields
$\{v,y,P_v,P_y\}$ are sufficiently smooth as functions of $r$
over the strictly positive axis. Then, any possible divergence
in the integrals over $r$ that appear in the metric must arise
either at infinity or around $r=0$. We also assume that the
inequality $2D>c_{\theta}^2\Omega$ (necessary for $w$ to be
real) is satisfied and that the value of $y$ when
$r\rightarrow\infty$ has already been set equal to zero by a
scaling of coordinates, fields and parameters \cite{me}.
However, we will not impose yet the vanishing of the limits
$v_0$ and $v_{\infty}$ of the field $v$ on the axis and at
infinity, respectively, although we will suppose that these
limits are, at least, finite. So, for the moment, we will allow
the presence in those regions of a dislocation in the $z$
direction.

In the rest of the paper, we will employ the following notation
\cite{di}. For any constant number $a$, the symbol
$f=\bar{o}(r^a)$ when $r\rightarrow 0$ indicates that $f$ is
much smaller than $r^a$ when one approaches the axis, so that
the limit of $ fr^{-a}$ vanishes at $r=0$. Likewise, the
notation $f=\bar{o}(r^a)$ when $r\rightarrow\infty$ means that
the limit of $fr^{-a}$ vanishes at infinity. On the other hand,
the symbol $f=o(g)$ when $r\rightarrow 0$ means that there
exists a strictly positive number $\epsilon$ such that the limit
of $ r^{-\epsilon}f/g$ is zero on the axis. In addition, the
symbol $f=o(g)$ when $r\rightarrow\infty$ implies that the limit
of $r^{\epsilon}f/g$ at infinity vanishes for a certain
$\epsilon>0$. In those occasions in which it is clear from the
context whether we are studying the behaviour of the solutions
close to the axis or, in contrast, at spacelike infinity, we
will employ the abbreviated notation $f=\bar{o}(r^a)$ and
$f=o(g)$, obviating the appearance of the limit $r\rightarrow 0$
or $r\rightarrow\infty$. More importantly, denoting by $\xi$ any
of the fields $\{v,y,P_v,P_y\}$, we will assume that the
behaviour at infinity and around $r=0$ is smooth enough to
guarantee that the condition $\xi=\bar{o}(r^a)$ when
$r\rightarrow \tilde{r}$ (with $\tilde{r}=0$ or $\infty$)
implies that $\partial_r\xi=\bar{o}(r^{a-1})$ and
$\partial_r^2\xi=\bar{o}(r^{a-2})$ at $\tilde{r}$. Similarly, if
$\xi=o(r^a)$ when $r\rightarrow
\tilde{r}$, we assume  that $\partial_r\xi=o(r^{a-1})$ and
$\partial_r^2\xi=o(r^{a-2})$ in the considered limit.

We will also write $f=O(r^a)$ if there exist positive constants
$M_1$, $M_2$, and $M_3$ such that $|fr^{-a}|<M_1$,
$|\partial_rfr^{1-a}|<M_2$, and $|\partial^2_rfr^{2-a}|<M_3$
close to the axis $r=0$. We will then say that the function $f$
is of the order or smaller than $r^a$ as $r\rightarrow 0$. In
practice, as we have commented, we will only consider solutions
whose fields are sufficiently smooth functions of $r$, including
the limit in which one reaches the symmetry axis. For this kind
of solutions, our definition of the symbol $f=O(r^a)$ amounts to
the existence of the limits of $fr^{-a}$, $\partial_rfr^{1-a}$,
and $\partial_r^2fr^{2-a}$ when $r\rightarrow 0$.

With the assumptions introduced above, it is possible to check
that our metric expressions are well defined for all values of
$r>0$ provided that $\bar{E}_{\infty}$ is finite and that, if
$c_{\theta}$ differs from zero, the integral $\Omega$ converges.
The first of these conditions, $\bar{E}_{\infty}<\infty$,
implies that $\bar{H}=\bar{o}(r^{-1})$ on the axis and at
infinity. Actually, a sufficient condition for
$\bar{E}_{\infty}$ to be finite is that $\bar{H}=o(r^{-1})$ in
the two considered regions. Although weaker conditions are
possible, we will restrict our discussion to this quite general
case from now on. Taking into account that $\bar{H}$ is the sum
of four non-negative factors, it is readily seen that our
condition can be equivalently expressed in the form
\begin{eqnarray}\label{conD}
&& \partial_rv=o(1),\hspace*{.8cm} \partial_ry=o(r^{-1}),\\
\label{conP}
&&P_v=o(r^{-1}),\hspace*{.5cm}P_y=o(1),
\end{eqnarray}
both when $r$ tends to zero and to infinity. For sufficiently
smooth fields $v$ and $y$ on the axis, one then obtains that,
around $r=0$,
\begin{equation}\label{conA}
v=v_0+o(r),\hspace*{.5cm}y=y_0+o(1).\end{equation} Here, $y_0$
is the limit of $y$ at $r=0$, which we suppose finite.
Similarly, far from the axis,
\begin{equation}\label{conI}
v=v_{\infty}+\bar{o}(1),\hspace*{.5cm}y=o(1).\end{equation}

In the case that $c_{\theta}$ differs from zero, one still has
to impose that $\Omega$ is finite. The convergence of the
integral at infinity is already ensured by the finiteness of
$\bar{E}_{\infty}$. The convergence at $r=0$, on the other hand,
can be seen to require that $\bar{H}=\bar{o}(r)$ on the axis,
and is satisfied, for instance, if the stronger condition
$\bar{H}=o(r)$ holds. Similar arguments to those presented above
show that this last condition implies that, when $r\rightarrow
0$,
\begin{equation}\label{conS}
v=v_0+o(r^2),\hspace*{.5cm} y=y_0+o(r),\hspace*{.5cm} P_v=o(1),
\hspace*{.5cm} P_y=o(r).\end{equation}
These conditions substitute for equations (\ref{conP}) and
(\ref{conA}) on the axis when the spacetime contains a
non-vanishing angular momentum.

Let us finally discuss the behaviour of the metric components
(with our choice of coordinates) in the limit $r\rightarrow 0$.
Although the axis is in principle excluded from our spacetime,
the possible singularities at $r=0$ are indeed quite weak,
allowing the metric to be well defined there. In the absence of
spin, $c_{\theta}=0$, it is straightforward to check that the
metric components are finite on the axis when conditions
(\ref{conP}) and (\ref{conA}) are satisfied. For spinning
solutions, on the other hand, the condition $v=v_0+o(r^2)$ when
$r$ is small, together with the convergence of $\Omega$, can be
seen to guarantee that $N^z-vN^{\theta}$ has a finite limit when
$r\rightarrow 0$. Then, the only metric component that can
diverge on the axis is the diagonal $t$ component. The
potentially divergent terms in that component come from the
contribution of $-e^y[e^{2w}\bar{N}^2-r^2(N^{\theta})^2]$. Given
the expression of $w$ and that $\bar{H}=o(r)$, it turns out
\cite{me} that this expression has nevertheless a finite limit
when $r\rightarrow 0$. Hence, the inclusion of spin does not
destroy the finiteness of the metric at $r=0$.

\section{Dynamical Stability}
\setcounter{equation}{0}

Since the metric expressions must be rigorously defined at all
instants of time, the boundary conditions on the fields must be
satisfied at every single moment and, therefore, be compatible
with the evolution dictated by equations
(\ref{vdot})-(\ref{Pydot}). Otherwise, the system would not
admit solutions that respect the conditions imposed at $r=0$ and
at infinity. In particular, this would imply that the metric
expressions diverge on the solutions of the model, so that they
would not lead to acceptable spacetimes. In this section, we
will analyse whether the conditions introduced in section 3 are
dynamically stable and, if the answer is in the negative,
replace them with stronger conditions that are preserved in
time.

Let us first study the asymptotic region far from the axis,
$r\rightarrow\infty$. Remembering that $e^{2w_{\infty}}$ and
$\bar{E}_{\infty}$ are positive and finite, it is not difficult
to check that equations (\ref{conP}) and (\ref{conI}), together
with our equations of motion, imply that
\begin{equation}
\dot{v}=o(1),\hspace*{.5cm}\dot{y}=o(r^{-1}),\hspace*{.5cm}
\dot{P}_v=\bar{o}(r^{-3}),\hspace*{.5cm}\dot{P}_y=o(r^{-1}).
\end{equation}
Then, our boundary conditions at infinity are stable, because
the time derivatives provide subdominant contributions, compared
to the leading terms in the fields. In particular, the behaviour
$\dot{v}=o(1)$ ensures that $v_{\infty}$ is time independent.
From now on, we will employ the notation $v_{\infty}^c$ to
remember the fact that this limit is constant.

The analysis of the stability around the symmetry axis $r=0$ is
much more involved. The solutions with and without spin must be
studied separately, because the factor $e^{-2w}$ that appears in
the equations of motion behaves in a different way: it has a
finite limit when $r\rightarrow 0$ if $c_{\theta}$ vanishes, but
diverges like $c_{\theta}^2/(4r^2)$ otherwise. In subsections
4.1 and 4.2 we will study, respectively, the families of
spacetimes with zero and non-vanishing angular momentum.

Before doing this, nevertheless, let us point out that there
exists at least an infinite family of solutions whose behaviour
around the axis is the same in the presence or absence of spin.
These solutions satisfy much stronger boundary conditions at
$r=0$ than those proposed in section 3. Namely, the fields
$v-v_0$, $y-y_0$, $P_v$ and $P_y$, as well as their derivatives
of any order with respect to $r$, decrease at $r=0$ faster than
any homogeneous polynomial of $r$. In other words, calling
$\{\psi\}\equiv\{v-v_0,y-y_0,P_v,P_y\}$, we have that the limit
of $r^{-a}\partial_r^m\psi$ vanish when $r\rightarrow 0$ for all
non-negative values of the integer numbers $a$ and $m$.
Examining the equations of motion (\ref{vdot})-(\ref{Pydot}),
one can see that these boundary conditions are indeed preserved
by the evolution. The reason is that the right-hand side of
those equations are given by terms in which we always find one
of our fields, or one of their derivatives, multiplied by a
factor that diverges, at most, like a negative power of $r$.
Therefore, all time derivatives turn out to vanish faster than
any positive power of $r$ on the axis. This proves the stability
and implies that the limits of $v$ and $y$ at $r=0$ (i.e. $v_0$
and $y_0$) must be time independent on those solutions. Note
also that the above set of spacetimes contains as a particular
subfamily the solutions in which one can find a neighbourhood
$r<r_0$ of the axis (with $r_0$ being a positive number) where
the momenta $P_v$ and $P_y$ vanish and the fields $v$ and $y$
are constant. The existence of this special, infinite family of
solutions was already noted in \cite{me}. Finally, let us
comment that the rapid decrease of the fields at $r=0$ in these
cylindrical spacetimes indicates a trivial interaction between
the vacuum degrees of freedom and the axial singularity.
Consequently, had we restricted our analysis to just this type
of solutions, the physical interest of our discussion would be
severely limited.

\subsection{Vanishing angular momentum}

We will now discuss the stability of the boundary conditions on
the axis when the angular momentum vanishes. In this subsection,
we will restrict our considerations to solutions that admit an
expansion in powers of $r$ around the symmetry axis. Obviously,
these do not include the solutions with rapid decrease of the
fields at $r=0$ commented above.

When $c_{\theta}$ vanish, the factor $e^{-2w}\bar{E}[r]$ that
appears on the right-hand side of all the equations of motion
reduces to the constant parameter $D$, greatly simplifying the
calculations. In addition, accepting the existence of power
series for the fields, the conditions (\ref{conP}) and
(\ref{conA}) translate into $v=v_0+O(r^2)$, $y=y_0+O(r)$,
$P_v=O(1)$ and $P_y=O(r)$. Employing this behaviour and the
dynamical equation for $v$, one concludes that, in fact,
$P_v=O(r)$ and the limit of $v$ must be constant. We will denote
this constant value by $v_0^c$. Finally, equation (\ref{Pydot})
and our conditions restrict the field $y$ to have the form
$y=y_0+O(r^2)$. The limit $y_0$, on the other hand, does not
need to be constant, but can vary in the evolution.

In conclusion, the dynamically stable boundary conditions on the
axis are
\begin{equation}\label{ANS}
v=v_0^c+O(r^2),\hspace*{.5cm}y=y_0+O(r^2),\hspace*{.5cm}
P_v=O(r),\hspace*{.5cm}P_y=O(r).\end{equation} In particular,
these conditions apply to cylindrical waves, case in which
$D=1$. Although the boundary conditions for these waves have
already been discussed in the literature \cite{api,roto}, we
have included their analysis for completeness. More importantly,
it seems that the results of \cite{roto} for waves with general
polarization contain some mistakes. Equations (\ref{conP}),
(\ref{conI}) and (\ref{ANS}) correct previous proposals and
provide the behaviour that must be imposed on the fields in the
asymptotic regions far and around the axis. For vanishing fields
$v$ and $P_v$, on the other hand, these conditions can be seen
to agree (modulo the supplementary assumption of expansions in
powers of $r^{-1}$ at infinity) with those imposed by Ashtekar
and Pierri for linearly polarized waves.

A more detailed analysis of the compatibility of the boundary
conditions and the dynamics shows that the power series of the
fields must be of the form
\begin{eqnarray}
&&v=v_0^c+\sum_{m=0}^{\infty}V_m r^{2m+2},\hspace*{.5cm}
P_v=\sum_{m=0}^{\infty}P_m r^{2m+1},\nonumber\\ \label{Tay}
&&y=y_0+\sum_{m=0}^{\infty}Y_m r^{2m+2},\hspace*{.5cm}
P_y=\sum_{m=0}^{\infty}Q_m r^{2m+1}.\end{eqnarray} The
coefficients of these series (except $v_0^c$) are functions of
the rescaled time $\tau=F_{\infty} t$ (where we have used that
$F_{\infty}$ is a constant of motion in the absence of spin). In
fact, let us suppose that we know, at all instants of $\tau$,
the value of $y_0$ and the coefficients $\{V_m,Y_m,P_m,Q_m\}$
for all $m\leq n$, $n$ being an integer. It is not difficult to
check that the power expansion of the equations of motion
provides then all the information needed to determine the
next-order approximation to the fields, i.e. the coefficients
with subindex equal to $n+1$. Moreover, the power series turn
out to be determined just by the functions of time $y_0$ and
$V_0$ and the constant $v_0^c$. Actually, once these
coefficients are known, the lowest-order contributions in the
dynamical equations for $v$ and $y$ allow one to obtain the
values of $P_0$ and $Q_0$, respectively. In addition, with
equation (\ref{Pydot}) and the knowledge of $y_0$, $V_0$ and
$Q_0$, one can fix $Y_0$. The iterative process outlined above
leads then to the determination of all other coefficients.

Finally, let us note that, instead of the functions of the
rescaled time $y_0$ and $V_0$, one can choose as degrees of
freedom in the power series the values of $y_0$ and of all the
sets of coefficients $\{V_m,Y_m,P_m,Q_m\}$ at a fixed, initial
time $\tau_0$. These values, together with $v_0^c$, completely
determine the fields at $\tau_0$. Given such initial data, the
equations of motion, which are first-order differential
equations, fix the evolution of the fields.

\subsection{Case with spin}

Let us now search for stable boundary conditions at $r=0$ when
$c_{\theta}$ does not vanish. Like in the previous subsection,
we will only consider solutions whose fields $\{v,y,P_v,P_y\}$
admit a power expansion in $r$ around the symmetry axis. Again,
this assumption does not hold in the family of spacetimes with a
rapid decrease of the fields discussed at the start of section
4. In addition, we will restrict our discussion to solutions in
which the field $v$ has a fixed constant limit at $r=0$, which
we will call $v_0^c$. In contradistinction with the situation
found for vanishing angular momentum, where the stability of the
boundary conditions requires that $v_0$ should be constant, a
heuristic analysis of the equations of motion seems to indicate
that now $\dot{v}_0$ may actually differ from zero. However, we
will concentrate our attention exclusively on solutions whose
spacelike helical structure in the vicinity of the axis
corresponds, at most, to a constant dislocation \cite{dislo}.
This includes, in particular, the case of a spinning cosmic
string \cite{djt, maz}, in which the dislocation is absent. The
possibility of $v_0$ being allowed to vary in time will be
discussed elsewhere.

With our hypotheses, conditions (\ref{conS}) for the convergence
of the integrals appearing in the metric expressions become
\begin{equation}\label{A1S}
v=v_0^c+O(r^3),\hspace*{.5cm}y=y_0+O(r^2),\hspace*{.5cm}P_v=O(r),
\hspace*{.5cm}P_y=O(r^2).\end{equation}
On the other hand, according to equation (\ref{w}), the factor
$e^{-2w}\bar{E}[r]$ that is present in all of the dynamical
equations diverges at $r=0$ like the inverse square of $r$ when
$c_{\theta}\neq 0$. Therefore, the condition that $P_v$ should
be at most of order $r$ is compatible with the equation of
motion for this momentum only if $v=v_0^c+O(r^4)$. With $v_0^c$
being a constant, the equation for $\dot{v}$ then requires that
$P_v$ must satisfy the stronger requirement $P_v=O(r^5)$.
Concerning the conditions on $y$ and $P_y$, we have, in fact,
two possibilities. For a generally time-dependent value of
$y_0$, equation (\ref{ydot}) imposes that $P_y=O(r^3)$. Then, it
turns out that the subdominant correction to $y_0$ must be at
least of order $r^6$. The reason is that, owing to the time
independence of the parameter $D$, the system must satisfy the
consistency condition $P_v\partial_rv+P_y\partial_ry=o(r^4)$, as
we will see in section 5. This requirement, together with the
stability of the condition on $P_y$, leads to the result
$y=y_0+O(r^6)$. The second possibility is that the value of
$y_0$ is a fixed constant, $y_0^c$. One would thus have
$y=y_0^c+O(r^2)$. In this case the equation of motion for $y$
imposes the condition $P_y=O(r^5)$.

We may express these two sets of stable boundary conditions in
the symbolic form
\begin{equation}\label{A2S}
v=v_0^c+O(r^4),\hspace*{.5cm}P_v=O(r^5),\hspace*{.5cm}
y=y_0^{(\kappa)}+O(r^{4+2\kappa}),\hspace*{.5cm}P_y=
O(r^{4-\kappa}),\end{equation} where the parameter $\kappa$ can
be equal to either plus or minus unity, and $y_0^{(1)}$
generally depends on time, whereas $y_0^{(-1)}$ is an
alternative notation for $y_0^c$. Note that the possibility
$\kappa=-1$ was not included in the discussions of \cite{me}.

With the above boundary conditions, one can see that, on
solutions to the equations of motion, the power expansions of
the fields must be of the following type
\begin{eqnarray}
&&v=v_0^c+\sum_{m=0}^{\infty}V_m r^{2m+4},\hspace*{1.2cm}
P_v=\sum_{m=0}^{\infty}P_m r^{2m+5},\nonumber\\ \label{Tas}
&&y=y_0^{(\kappa)}+\sum_{m=0}^{\infty}Y_m
r^{2m+4+2\kappa},\hspace*{.5cm} P_y=\sum_{m=0}^{\infty}Q_m
r^{2m+4-\kappa}.\end{eqnarray} Except $v_0^c$ and
$y_0^{(-1)}=y_0^c$, which are constants, all the coefficients in
these series depend, in principle, on the modified time
coordinate $\tau=\int_0^t F_\infty(\bar{t})d\bar{t}$. This
redefinition of time absorbs the common factor $F_{\infty}$ that
appears in the dynamical equations.

Actually, all the coefficients in these series can be found if
one knows, at all instants of $\tau$, the values of $v_0^c$,
$V_0$, and $y_0^{(\kappa)}$, as well as the value of $Y_0$ when
$\kappa=-1$. The proof of this statement is sketched in appendix
A. The main difference with respect to the situation described
when the angular momentum vanishes is that now the coefficients
of the power expansion are not all functionally independent;
instead, they satisfy relations which do not involve time
derivatives. This issue is also discussed in appendix A. As a
result of such functional relations, one can, in fact, prove by
induction that, at a generic initial time $\tau_0$, the series
(\ref{Tas}) can be completely determined if one knows at that
moment just the values of $v_0^c$, $y_{0}^{(\kappa)}$ and all
the sets of the form $\{V_{3n},Q_{3n},P_{3n},Y_{3n}\}$, where
$n$ is any non-negative integer. Note that, since the dynamics
is dictated by first-order differential equations, the commented
collection of coefficients, evaluated at $\tau_0$, provide then
all the information needed to fix the power expansion of the
fields around $r=0$ at all instants of time.

From these comments, it is also clear that a possible procedure
to construct admissible solutions is the following. At a certain
initial time, determine the initial values of the fields in a
region around $r=0$ by fixing the collection of independent
coefficients given above. Try then to analytically continue such
initial values to the whole positive semiaxis. If the
continuation is possible and satisfies the boundary conditions
(\ref{conP}) and (\ref{conI}), use such initial data, together
with the equations of motion, to arrive at a physical solution.
Otherwise, employ as initial data the result of a smooth
matching between the initial values obtained around $r=0$ and
any initial fields in the region far from the axis that satisfy
the boundary conditions at infinity.

\section{Consistency of the formalism}
\setcounter{equation}{0}

In the previous section, we have proved that there exist
boundary conditions that ensure that the metric expressions,
which had been obtained by means of formal integrations, are
meaningful at all instants of time on the solutions of the
system. In order to prove that such solutions lead, in fact, to
physically acceptable spacetimes with a rigorously defined
energy density, we want to show now that the introduced boundary
conditions guarantee also that the dynamics is fully consistent
and that the system possesses a well defined (reduced)
Hamiltonian formalism. Let us remind that, in the asymptotic
region $r\rightarrow\infty$, the boundary conditions are given
by equations (\ref{conP}) and (\ref{conI}), where the limit of
$v$ is a constant (i.e. $v_{\infty}=v_{\infty}^c$). On the other
hand, close to the axis $r=0$, the behaviour of the fields is
dictated by equation (\ref{ANS}) if the angular momentum
vanishes, and by equation (\ref{A2S}) if $c_{\theta}$ differs
from zero. Remember also that, in this last case, the constant
$\kappa$ can adopt the values $\pm1$. In addition, it is worth
noting that, although this behaviour on the axis was deduced
assuming that the basic fields admit power expansions in $r$,
the conditions apply as well to the set of solutions with rapid
decrease of the fields at $r=0$ discussed in section 4. So, for
all of the considered solutions, it will suffice to prove the
consistency of the Hamiltonian dynamics when requirements
(\ref{ANS}) or (\ref{A2S}) are satisfied.

We will first prove that the dynamics is compatible with the
time independence of the parameter $D$, assuming that the
equations of motion remain valid in the limit $r\rightarrow 0$.
Note that, if this compatibility could not be reached, one would
be forced either to admit that the system does not possess
physically acceptable solutions or to try and introduce rather
artificial sources on the symmetry axis that could account for
the constancy of $D$.

The constant $D$ determines the subleading term in $e^{2w}$
around the axis $r=0$ in the presence of spin, and the leading
contribution when $c_{\theta}$ vanishes. In the case with
non-zero angular momentum, using that $\bar{H}=o(r)$ for
$r\rightarrow 0$, one can check
\begin{equation}\label{D4}
e^{2w}=\frac{4}{c_{\theta}^2}r^2-\frac{16D}{c_{\theta}^4}r^4
+o(r^4).\end{equation} Similarly, when there is no angular
momentum, the condition $\bar{H}=o(r^{-1})$ ensures that
$e^{2w}=1/D+o(1)$. Therefore, the hypothesis that $D$ is
constant is compatible with the dynamics if and only if
$\partial_t(e^{2w})=o(r^4)$ when $c_{\theta}\neq 0$, and
$\partial_t(e^{2w})=o(1)$ when $c_{\theta}$ vanishes. On the
other hand,the equations of motion for our system, together with
the expression of $e^{2w}$, lead to \cite{me}
\begin{equation}\label{wdot}
\partial_t(e^{2w})=2F_{\infty}\bar{E}[r]\left(P_v\partial_r
v+P_y\partial_ry\right).\end{equation} Since $\bar{E}[0]=1$, we
then conclude that the term $P_v\partial_r v+P_y\partial_r y$
must be of the form $o(r^4)$ around the axis when $c_{\theta}$
differs from zero, and of the form $o(1)$ in the absence of
spin. It is then easy to check that the respective boundary
conditions (\ref{A2S}) and (\ref{ANS}) guarantee this
requirement.

In order to present a well defined Hamiltonian dynamics, our
family of cylindrical spacetimes must not only possess a real
and finite reduced Hamiltonian $H_R$, but this Hamiltonian must
also be differentiable on phase space. Otherwise, $H_R$ would
not generate a true canonical transformation and, therefore, a
valid time evolution \cite{ama}. Note that, if that happened to
be the case, the system would be missing an acceptable notion of
energy density. We have already commented that, once the
integrals $\bar{E}_{\infty}$ and $\Omega$ that appear in
equation (\ref{Hom}) are known to converge, the reality and
finiteness of $H_R$ amount just to the condition
$2D>c_{\theta}^2\Omega$. This inequality is preserved in the
evolution, and can hence be regarded as a mere restriction on
the initial values of the fields \cite{me}. Moreover, there
always exists a non-empty region of phase space where this
inequality is satisfied. On the other hand, it is proved in
appendix B that our boundary conditions guarantee also that the
Hamiltonian $H_R$ is, in fact, differentiable. Hence, our
conditions, together with the restriction
$2D>c_{\theta}^2\Omega$ on the initial values, ensure that the
Hamiltonian dynamics is well defined and that there exists a
meaningful constant of motion that provides the energy density.

\section{Metric at spacelike infinity}
\setcounter{equation}{0}

Once we have determined the boundary conditions that must be
satisfied by the basic fields of the system, we can discuss the
behaviour of the metric close to the axis $r=0$ and at spacelike
infinity, $r\gg 1, |t|$. The knowledge of this behaviour will be
essential to determine the geometrical properties of our
solutions and reach a well founded physical interpretation of
the vacuum spacetimes under consideration. Let us first analyse
the region at spacelike infinity.

Taking into account that when $r\rightarrow\infty$ the behaviour
of the fields is governed, for all values of the spin parameter
$c_{\theta}$, by conditions (\ref{conP}) and (\ref{conI}), with
$v_{\infty}=v_{\infty}^c$ being a constant, a trivial
calculation shows
\begin{eqnarray}
&&\bar{E}[r]=\bar{E}_{\infty} [1+o(1)],\hspace*{.8cm}
e^{2w}=e^{2w_{\infty}}+o(1),\nonumber\\
&&N^{\theta}=\frac{c_{\theta}}{2r^2}e^{w_{\infty}}+o(r^{-2}),
\hspace*{.5cm}N^z-vN^{\theta}=\bar{o}(r^{-2}).\end{eqnarray}
The cylindrical metric adopts then the asymptotic form
\begin{eqnarray}
&ds^2=&[1+o(1)]\left[-\left(dt-e^{w_{\infty}}\frac{c_{\theta}}{2}d
\theta\right)^2+r^2d\theta^2+e^{2w_{\infty}}dr^2\right]\nonumber\\
\label{asyI}
&&+[1+o(1)]\left\{dz+\bar{o}(r^{-2})dt-[v_{\infty}^c+\bar{o}(1)]d
\theta\right\}^2.
\end{eqnarray}
Disregarding the possible existence of a constant dislocation in
the $z$ direction (i.e. taking $v_{\infty}^c=0$) and neglecting
contributions of the form $o(1)$ in the metric, as well as terms
of the type $o(r^2)$ and $\bar{o}(1)$ in the diagonal $\theta$
component and the $z\theta$ component, respectively, we obtain
precisely the metric that would be created by a spinning cosmic
string. Moreover, the density of angular momentum of the studied
cylindrical solution, $c_{\theta}/2$, coincides with the spin
(per unit length) of the string that would produce the
approximate metric. In addition, the reduced Hamiltonian of our
spacetime equals the energy density of the string, namely,
$1-e^{-w_{\infty}}$. Obviously, the deficit angle at spacelike
infinity is simply this energy density multiplied by a factor of
$2\pi$.

The resemblance in the analysed asymptotic region to the
exterior metric of a spinning string is considerably enhanced
when one restricts one's attention to solutions with
\begin{equation}\label{asyC}
v=\bar{o}(r^{-2}),\hspace*{.5cm}y=\bar{o}(r^{-2}),
\hspace*{.5cm}P_v=\bar{o}(r^{-3}),\hspace*{.5cm}
P_y=\bar{o}(r^{-1}).\end{equation} Note that this behaviour
guarantees that conditions (\ref{conP}) and (\ref{conI}) are
satisfied. A particular subfamily of spacetimes in which these
asymptotic restrictions hold is that formed by all solutions
with fields of compact support, namely, (smooth) solutions whose
fields $v$, $y$, $P_v$ and $P_y$ vanish in the region $r\geq
r_1$, with $r_1$ being a positive number. From the equations of
motion of our system, it is not difficult to check that the
compact support of these fields is respected in the evolution,
so that the noted kind of solutions exist. Employing conditions
(\ref{asyC}), a detailed calculation leads to the following
metric at spacelike infinity:
\begin{eqnarray}
&ds^2=&[1+\bar{o}(\rho^{-2})]\left[-\left(dt-e^{w_{\infty}}
\frac{c_{\theta}}{2}d\theta\right)^2+\rho^2d\theta^2+
e^{2w_{\infty}}d\rho^2+dz^2\right]\nonumber\\
&&+\bar{o}(\rho^{-2}) dzd\theta+\bar{o}(\rho^{-4})dtdz.
\end{eqnarray}
Here $\rho^2=r^2+e^{2w_{\infty}}c_{\theta}^2/4$ is a new radial
coordinate. So we see that the approximation in the asymptotic
region to the exterior metric of a (possibly) spinning string
has been improved, with respect to the general case, to the
level $\bar{o}(1)$ in the diagonal $\theta$ component and at
least to the level $\bar{o}(\rho^{-2})=\bar{o}(r^{-2})$ in the
rest of the metric components.

\section{Metric near the axis}
\setcounter{equation}{0}
\subsection{Vanishing angular momentum}

Since the boundary conditions at $r=0$ are different for
solutions with and without angular momentum, the corresponding
behaviour of the metric must be studied separately. Let us first
analyse the case in which the axis is not endowed with spin. The
conditions (\ref{ANS}) on the fields imply that
$\bar{E}[r]=1+O(r^2)$, and it is then straightforward to see
that
\begin{equation} \label{asyA}
ds^2=[1+O(r^2)]e^{y_0}\left[-d\bar{t}^2+r^2d\theta^2+\frac{dr^2}{D}
+e^{-2y_0}d\bar{z}^2\right]+O(r^2)d\bar{z}d\theta,
\end{equation}
where $\bar{z}=z-v_0^c\theta$ and
$\bar{t}=t/\sqrt{\bar{E}_{\infty}}$ is a rescaled time. In
defining this coordinate, we have employed that
$\bar{E}_{\infty}=De^{2w_{\infty}}$ is a constant of motion when
$c_{\theta}$ vanishes. Neglecting contributions or order $r^4$
to the diagonal $\theta$ component and of order $r^2$ to the
rest of the metric elements, we see that, near the axis, the
metric describes the exterior of a straight cosmic string,
except for the possible periodic structure in the $\bar{z}$
coordinate and the allowed time dependence of the factor
$e^{y_0}$, which remains constant for a static string.

Several comments are in order at this point. First, let us
remark that the time coordinates used in our approximations near
the axis and at spacelike infinity differ by a scaling that
depends on the considered solution. It is not difficult to see
that this scaling reduces to the identity only for the flat
solution with vanishing fields $v-v_0^c$, $y$, $P_y$ and $P_v$.
This solution is possible only if $v_0^c=v_{\infty}^c$ and
$y_0=0$.

On the other hand, the periodicity of $\theta$ leads to the
identification of points $(\theta,\bar{z})$ and
$(\theta+2\pi,\bar{z}-2\pi v_0^c)$. To avoid this spacelike
helical structure, we will assume in the following that $v_0^c$
vanishes. Let us also note that there exist solutions in which
$y_0$ is actually constant. For instance, this occurs in the
solutions with fields that decrease rapidly at $r=0$, as
discussed in section 4. Furthermore, we proved in subsection 4.1
that the time-dependent coefficients $y_0$ and $V_0$ could be
considered as the true degrees of freedom contained in the
expansions (\ref{Tay}) around the axis. Therefore, the
possibility that $y_0$ is set equal to a constant is indeed
available.

In addition, it is known that (at least when the angular
momentum vanishes) cylindrical gravity in a vacuum can be
reduced to three dimensions by employing the existence of a
translational Killing vector \cite{kra,ama}. This reduction
leads to three-dimensional gravity with axial symmetry coupled
to two scalar fields, namely $v$ and $y$ \cite{anme}. From this
point of view, the metric in three dimensions associated with
our spacetimes in the vicinity of the axis is
$ds^2=-d\bar{t}^2+r^2d\theta^2+dr^2/D$, modulo relative
corrections of the order of $r^2$ or smaller. This is exactly
the 3-metric created by a point particle of mass $m=1-\sqrt{D}$
\cite{djt}. With this perspective, the time variation displayed
by $y_0$ in equation (\ref{asyA}) can be understood just as a
time-dependent lift of the metric from three to four dimensions.
This time dependence is generally necessary to compensate the
radial variation of the fields around the axis, so as to finally
arrive at a vacuum solution for $r>0$.

We also note that the metric (\ref{asyA}) presents a deficit
angle around the axis, a property that is characteristic of
stringy defects. Moreover, this deficit angle is independent of
time, namely $2\pi(1-\sqrt{D})$, as it would correspond to a
static string with linear energy density equal to $m$, i.e. the
mass of the point particle obtained in the three-dimensional
version of the system. Thus, we see that, like in the case of a
straight string, the effect of the singularity on the axis is to
introduce a constant deficit when $D\neq 1$. Note that, in order
for this deficit to be positive, one must restrict the parameter
$D$ to be smaller than unity. For $D=1$, on the other hand, the
spacetime is completely regular \cite{me} and the considered
solutions are purely gravitational waves. Finally, it is worth
commenting that the approximations of the metric given in
equations (\ref{asyI}) and (\ref{asyA}) can be checked to
describe the behaviour of the solutions with quasi-regular axis
analysed in Refs. \cite{xan,etso}, as well as those contained in
the soliton spacetimes of class A2 and B2 constructed by Garriga
and Verdaguer \cite{vega}. In agreement with the above
discussion, such solutions have been interpreted as representing
the interaction of cylindrical waves with a cosmic string.

\subsection{Case with spin}

When $c_{\theta}\neq 0$, there exist two acceptable sets of
boundary conditions (assuming a constant value for $v$ at
$r=0$), each of them leading to a different metric behaviour. We
will first consider the case in which the field $y$ may have a
time-dependent limit on the axis. The corresponding boundary
conditions are given by equation (\ref{A2S}) with $\kappa=1$.
Remember that these conditions apply as well to solutions with
fields that decrease rapidly at $r=0$, even though they were
originally deduced for fields that admit an expansion in powers
of $r$. One can then easily see that
\begin{eqnarray}
&& e^{-2w}=\frac{c_{\theta}^2}{4r^2}\left[1+\frac{r^2}{A^2}+
O(r^6)\right],\hspace*{.8cm}\bar{E}[r]=1+O(r^{6}),\nonumber\\
&&N^{\theta}=\frac{c_{\theta}}{2r^2}F_{\infty}[1+2\Omega r^2+
O(r^6)],\hspace*{.5cm}
N^z-vN^{\theta}=N_0^z+O(r^2).\end{eqnarray} Here, $\Omega$ is
the convergent integral defined in equation (\ref{Hom}) and
\begin{equation} \label{A}N_0^z=c_{\theta}F_{\infty}\int_0^
{\infty}\frac{dr}{r^3}(v-v_0^c)\bar{E}[r],\hspace*{.5cm}
A=\frac{c_{\theta}}{2\sqrt{D}}.\end{equation} Note that $N^z_0$
is finite, given our boundary conditions. Introducing then the
coordinates
\begin{eqnarray}\label{newco}
&&\bar{t}=\sqrt{D}\int_0^t
d\tilde{t}\,F_{\infty}(\tilde{t}),\hspace*{.5cm}\bar{\theta}=
\theta+c_{\theta}\int_0^td\tilde{t}\,\Omega(\tilde{t})F_{\infty}
(\tilde{t}),\nonumber\\ &&\rho^2=r^2+A^2,\hspace*{1.7cm}
\bar{z}=z-v_0^c\theta+\int_0^t d\tilde{t}\,N_0^z(\tilde{t}),
\end{eqnarray} the
metric around the axis $r=0$ can be expressed in the form
\begin{eqnarray}\label{asy6}
&ds^2=&[1+O(r^6)]e^{y_0^{(1)}}\left[-\left(d\bar{t}-Ad\bar
{\theta}\right)^2+\rho^2d\bar{\theta}^2+\frac{d\rho^2}
{D}+e^{-2y_0^{(1)}}d\bar{z}^2\right]\nonumber\\
&&+O(r^{2})d\bar{t}d\bar{z}+O(r^{4})d\bar{z}d\bar{\theta}
+O(r^4)(d\bar{t})^2.\end{eqnarray}

In the definition of our new coordinates, we have taken into
account that the quantities $F_{\infty}$, $N_0^z$ and $\Omega$
may implicitly depend on time when $c_{\theta}\neq 0$. Like in
the absence of angular momentum, the time and angular
coordinates of the expansions around the axis and at spacelike
infinity [see equation (\ref{asyI})] differ for all but the flat
solution, which has vanishing momenta $P_v$ and $P_y$ and
constant fields $v=v_0^c$ and $y=0$. Concerning the axial
coordinate $\bar{z}$, we see that the periodicity of $\theta$
introduces again non-trivial identifications of points in the
sections of constant time unless $v_0^c$ vanishes. We will thus
restrict our discussion to the case $v_0^c=0$ from now on. The
coordinates $z$ and $\bar{z}$ can then be seen to coincide on
the flat solution; otherwise they generally differ. Finally,
note that the new coordinate $\rho$ is defined only over the
semiaxis $(|A|,\infty)$, because, in our spacetime, $r$ must be
positive. In principle, however, one could try and make an
analytic extension to values of $\rho$ smaller than $|A|$. This
would correspond to imaginary values of $r$. It is important to
remark that, in any case, such an extension would give rise to
the appearance of CTCs because, from equation (\ref{asy6}) and
at least in the region with $\rho$ smaller but close to $|A|$,
the diagonal $\bar{\theta}$ component of the metric would then
be negative, indicating the presence of a timelike vector field
with closed orbits.

In terms of $\rho$, the expression $O(r^{2a})$ can be rewritten
in the form $O([\rho-|A|]^a)$ for any real number $a$. On the
other hand, disregarding the small $\bar{t}\bar{z}$ and
$\bar{z}\bar{\theta}$ components, the contributions of the type
$O(r^4)$ to the diagonal $\bar{t}$ component, and all relative
corrections to the metric of the order of $r^6$, we arrive at a
metric near the section $\rho=|A|$ that describes the
gravitational field created by a spinning string, except for the
fact that now $y_0$ may depend on time. Of course, this time
dependence does not show up in the solutions with rapidly
decreasing fields, since $y_0$ is then a constant. In addition,
from our discussion in subsection 4.2, where we showed that the
function $y_0(t)$ could be considered as one of the degrees of
freedom contained in the series (\ref{Tas}), it follows that
there exist solutions with power series at $r=0$ in which $y_0$
remains constant in time.

In the general case in which $y_0$ depends on time, a line of
reasoning similar to that presented for spacetimes with
vanishing momenta allows us to regard the approximation of the
metric around $\rho=|A|$ as a time-dependent lift to four
dimensions of the three-dimensional metric produced by a
rotating point particle with mass \cite{djt}. In addition, it is
worth noting that the approximated metric again presents a
deficit angle that is constant. In this spirit, the arguments
given in subsection 7.1 support the interpretation of the metric
in the analysed region as that associated with a spinning string
interacting with an environment of gravitational waves. This
spinning string is characterized by two parameters, namely, the
corresponding density of angular momentum $c_{\theta}/2$ and the
deficit angle $2\pi(1-\sqrt{D})$. Actually, this value of the
deficit is not modified by the presence of spin, since it takes
the same expression when $c_{\theta}$ vanishes.

To end this section, we will discuss the behaviour of the metric
around the axis $r=0$ when the limit of $y$ is a fixed constant
$y_0^c$, so that the boundary conditions are given by equation
(\ref{A2S}) with $\kappa=-1$. These imply
\begin{eqnarray}
&& e^{-2w}=\frac{c_{\theta}^2}{4r^2}
\left[1+\frac{r^2}{A^2}+O(r^4)\right],\hspace*{.8cm}
\bar{E}[r]=1+O(r^{4}),\nonumber\\
&&N^{\theta}=\frac{c_{\theta}}{2r^2}F_{\infty}[1+2\Omega r^2+
O(r^4)],\hspace*{.5cm}
N^z-vN^{\theta}=N_0^z+O(r^2).\end{eqnarray} Then, employing
equation (\ref{newco}), we can write the metric in the form
\begin{eqnarray}\label{asy4}
&ds^2=&[1+O(r^4)]e^{y}\left[-\left(d\bar{t}-Ad\bar{\theta}
\right)^2+\rho^2d\bar{\theta}^2+\frac{d\rho^2}
{D}\right]+e^{-y}d\bar{z}^2\nonumber\\ &&+
O(r^{2})d\bar{t}d\bar{z}+O(r^{4})d\bar{z}d\bar{\theta}.
\end{eqnarray}
Here, $e^y=e^{y_0^c}+O(r^2)$. Note that, around $\rho=|A|$, the
metric presents, in fact, a constant deficit angle. The
corrections to this deficit are at most of the order of $r^4$,
even if the subdominant contributions to $y_0^c$ can be of the
type $O(r^2)$. Like in the case $\kappa=1$, we will restrict our
attention to the possibility that $v_0^c$ vanishes. Since
$y_0^c$ is a constant, it is then clear that the metric in the
region $0<\rho-|A|\ll 1$ can be interpreted as that originated
by a string with spin equal to $c_{\theta}/2$ and a conical
deficit given by $2\pi(1-\sqrt{D})$.

\section{Discussion and conclusions}

We have considered the most general cylindrical solution to the
Einstein equations in vacuo that, in principle, does not contain
the symmetry axis $r=0$. This axis, which may be singular, has
been allowed to possess spin. However, to arrive at metric
expressions that are well defined in the whole region $r>0$
\cite{me}, we have supposed that the linear momentum in the axis
direction vanishes. These spacetimes were analysed in a recent
work \cite{me}, where a gauge-fixing procedure that removes all
of the gravitational constraints was introduced. The resulting
reduced model can be described by four fields on phase space,
$\{v,y,P_v,P_y\}$, whose dynamics is generated, at least
formally, by a reduced Hamiltonian $H_R$. The singularity on the
axis can be characterized by two parameters: a real constant
$c_{\theta}$ that provides (twice) the density of angular
momentum and a positive number $D$ that determines the behaviour
of the purely radial component of the metric around $r=0$. The
latter of these parameters should be a constant in order for the
dynamics to be consistent and the system to possess a conserved
energy density \cite{me}, given by the value of $H_R$.

We have first searched for boundary conditions which ensure that
the reduced system obtained after completing the gauge fixing is
well defined. These conditions describe the behaviour of the
basic fields $\{v,y,P_v,P_y\}$ at spacelike infinity, $r\gg
1,|t|$, and near the axis $r=0$. More explicitly, the
consistency of the system implies the following requirements.
First, all the metric expressions which have been found by means
of formal integrations must be meaningful. Second, in order for
the reduced Hamiltonian to generate a well defined dynamics and
provide the energy density of the system, $H_R$ must be real,
finite and differentiable on phase space. Third, these
requirements must be satisfied at all instants of time and,
therefore, the boundary conditions must be dynamically stable.
Finally, assuming that the equation of motion for the metric
function $w$ (that appears in the diagonal radial component)
remains valid in the limit $r\rightarrow 0$, one has to check
that the value of the parameter $D$ is, in fact, preserved in
the evolution. The existence of boundary conditions that satisfy
the above requirements is most fundamental; otherwise, the
reduced dynamics would not be consistent and there would not
exist physically acceptable (non-trivial) solutions in
cylindrical vacuum gravity endowed with a non-vanishing angular
momentum.

In our discussion, we have supposed that our basic fields are
smooth over the whole semi-axis $r>0$ for all possible values of
the time coordinate $t$. In addition, denoting by $\xi$ any of
these fields, we have assumed that both the requirements
$\xi=o(r^a)$ and $\xi=\bar{o}(r^a)$, either at $r=0$ or at
infinity, automatically imply that the derivatives
$\partial_r\xi$ and $\partial_r^2\xi$ display a behaviour
similar to that of the field $\xi$, but with the exponent $a$
replaced with $a-1$ and $a-2$, respectively. With this
assumption, the boundary conditions at spacelike infinity turn
out to be given by equations (\ref{conP}) and (\ref{conI}). In
the latter of these equations, $v_{\infty}$ (i.e. the asymptotic
limit of $v$) is a fixed number that reflects the possible
existence of a constant dislocation in the $z$ direction. To
avoid the appearance of an asymptotic spacelike helical
structure \cite{dislo}, one only has to make $v_{\infty}=0$.

The boundary conditions that must be introduced on the axis
$r=0$, on the other hand, differ for solutions with or without
angular momentum. When $c_{\theta}$ vanishes, the conditions are
given by equation (\ref{ANS}). Again, to prevent the existence
of a screw dislocation in the axis direction \cite{dislo}, one
must demand that $v$ vanishes at $r=0$, i.e. $v_0^c=0$. For
non-zero spin, there exists more than one set of admissible
boundary conditions. We have only considered in detail the case
in which the limit of $v$ on the axis $r=0$ is constant, thus
corresponding to a fixed dislocation. Then, the behaviour of the
fields depend on whether the limit of $y$ is also fixed and time
independent or, in contrast, is allowed to vary in time. In both
situations, the boundary conditions can be written in the
symbolic form (\ref{A2S}), where $\kappa=-1$ for the first of
the considered possibilities ($y_0^{(-1)}$ is constant) and
$\kappa=1$ otherwise.

By means of a careful analysis of the equations of motion and
the Hamiltonian, we have proved that these boundary conditions
are stable and that, together with the requirement
$2D>c_{\theta}^2\Omega$ (which restricts the admissible initial
data for the basic fields), guarantee that the metric and
Hamiltonian dynamics are rigorously defined. We have also
checked that the evolution is compatible with the constancy of
$D$. So all consistency conditions are satisfied. In addition,
employing the boundary conditions, we have been able to
calculate the approximate form of the metric in the region close
to the axis $r=0$ and at spacelike infinity. The physical
picture that arises from this analysis is the following.

In the region near $r=0$ (and assuming that $v_0^c=0$), the
metric describes a `rotating' conical geometry, with constant
deficit angle and a generally non-vanishing angular momentum.
This approximate metric corresponds to the lift from three to
four dimensions of the metric produced by a point particle with
constant mass and spin given by $1-\sqrt{D}$ and $c_{\theta}/2$.
The lift is in general time dependent, to account for the
possible variation of the fields around $r=0$ and yield a
solution to the Einstein equations at all points of the
spacetime. The 4-metric can thus be regarded as that caused, in
the analysed region, by a spinning cosmic string with linear
density of energy and angular momentum determined by the mass
and spin parameters of the analogous particle in three
dimensions. Note that, in order for this energy density to be
non-negative, one must restrict the parameter $D$ to be equal or
smaller than unity. It is also worth remarking that the time and
cylindrical coordinates that are naturally associated with this
spinning string do not coincide with those selected in our gauge
fixing, which are specially adapted to the asymptotic region far
from the axis. The relation between both sets of coordinates is
given by equation (\ref{newco}) (with $v_0^c=0$). This relation
continues to be valid even if $c_{\theta}$ vanishes [see
equation (\ref{asyI})].

Of particular importance is the change of radial coordinate in
the presence of spin. This change is precisely that which would
remove the region with CTCs from the exterior of a spinning
cosmic string, mapping the resulting spacetime to the sector
$r>0$. This opens the possibility of analytically continuing our
spacetime to the region of imaginary values of $r$ (namely, to
$\rho<|A|$) at the price of introducing CTCs. Note that, in our
original spacetime, the existence of CTCs is actually precluded.
An infinite family of solutions in which the continuation to
imaginary values of $r$ can be straightforwardly performed is
that with fields of rapid decrease at $r=0$, considered in the
beginning of section 4. It is clear that the metric of any of
these spacetimes can be matched smoothly at $r=0$ with the
metric of a spinning cosmic string describing the region
$\rho\leq|A|$, providing in this way an extended solution that
contains timelike orbits generated by $\partial_{\theta}$.

Since our spacetimes are the most general solution to the
Einstein equations in a vacuum for $r>0$, the only field content
outside the axis $r=0$ is that corresponding to gravitational
waves. Consequently, the studied spacetimes represent an
ensemble of gravitational waves surrounding a singular axis that
is characterized by its spin density and by the constant deficit
angle detected in its vicinity. These are precisely the effects
that a spinning string would produce. In this sense, the
analysed spacetimes can be regarded as describing the most
general interaction that is permitted in general relativity
between cylindrical waves and strings with constant density of
energy and spin.

This interpretation is in accordance with the asymptotic form of
the metric (\ref{asyI}) at spatial infinity. It is known that,
in the region $r\gg 1,|t|$, a cylindrical wave causes a deficit
angle that is proportional to the total energy density contained
in the wave; namely, the deficit is
$2\pi(1-1/\sqrt{\bar{E}_{\infty}})$ \cite{roto}. In addition,
the wave does not carry angular momentum. Therefore, in the case
that the gravitational wave surrounds a cosmic string, the
deficit angle at spatial infinity should be produced by the
combined effect of both phenomena, and the angular momentum
should be originated in the string. So, for a cylindrical wave
interacting with a spinning string, one would expect the
asymptotic metric at spacelike infinity to describe a conical
geometry with a spin parameter equal to that of the string, but
with a different constant deficit angle. This is, in fact, the
result that we have obtained.

Let us analyse in more detail the relation between the deficit
angles encountered at $r=0$ and at infinity: $2\pi(1-\sqrt{D})$
and $2\pi(1-e^{-w_{\infty}})$, respectively. In the absence of
spin, the relation is relatively simple; at spacelike infinity,
one only has to divide the parameter $D$ by $\bar{E}_{\infty}$,
a constant of motion that determines the energy of the wave.
This quotient can be interpreted in terms of the associated
$C$-energy \cite{tho} as an addition of energies. In the
spinning case, on the other hand, the relation is much more
complicated. The quantity $\bar{E}_{\infty}$ is not a constant
of motion anymore, and the gravitational wave does not possess a
preserved energy density by its own (since the $C$-energy is not
conserved). Moreover, the existence of angular momentum
introduces corrections to the deficit angle that depend on the
gravitational fields not just through the functional
$\bar{E}_{\infty}$, but also via $\Omega$. The spin decreases
the effective value of $D$ by an amount of
$c_{\theta}^2\Omega/2$. The resulting deficit is, apart from the
usual factor of $2\pi$, equal to the total energy density of the
system, given in equation (\ref{Hom}) and which is again
conserved. Furthermore, as we commented in section 2, the
positivity of $\bar{H}$ on phase space implies that $H_R\geq
1-\sqrt{D}$. Consequently, in absolutely all of the spacetimes
that we have considered, the deficit angle at infinity is
greater than or equal to that around the axis $r=0$. This is a
general prediction that, in principle, could be verified
experimentally, had we the possibility of making measurements in
cylindrical gravitational systems containing a stationary
stringy defect or, alternatively, in systems that could mimic
the dynamics of the gravitational field in this situation.

Our result can also be rephrased by saying that, although the
energy of the composite system is not the sum of the energies
corresponding to the spinning string and the cylindrical wave,
the presence of a gravitational wave always results in an
increase of the total energy. It is not difficult to check that
the only situation in which the two studied deficit angles
coincide is when the fields $v$ and $y$ are constant and the
momenta $P_v$ and $P_y$ vanish all over the spacetime, i.e. for
the flat solution which (modulo a constant dislocation, given by
the fixed value of $v$) describes the vacuum region, free of
CTCs, in the exterior of a string with mass and spin densities
equal to $1-\sqrt{D}$ and $c_{\theta}/2$.

It is worth commenting that, although the degrees of freedom of
our family of spacetimes are those corresponding to purely
gravitational cylindrical waves (namely, the fields $v$ and $y$
and their canonical momenta), the equations of motion that
dictate the evolution in configuration space (i.e. in terms of
$v$ and $y$) are not truly hyperbolic partial differential
equations, except in the absence of spin. The reason is the
appearance of the factor $e^{-2w}\bar{E}[r]$ in all of the
dynamical equations (\ref{vdot})-(\ref{Pydot}). This factor has
a local dependence on the fields if and only if $c_{\theta}$
vanishes, in which case it reduces to the constant $D$. In other
words, the existence of angular momentum introduces, via the
process of gauge fixing \cite{me}, a high non-locality in the
dynamics of cylindrical gravity.

Finally, in order to confirm the interpretation that we have put
forward for our family of spacetimes as cylindrical waves
interacting with spinning strings, it would be interesting to
analyse the Riemann tensor of the solutions, paying a particular
attention to the contributions that, in the form of
distributions concentrated on line sources, could account for
the appearance of the axial singularity \cite{djt,ls}. In
addition, one might also study the behaviour of the Riemann
tensor at null infinity, determining the radiative content of
the gravitational field in this region (as it was done, e.g., in
Refs. \cite{etso,vega}). These issues will be the subject of
future research.

\section*{Acknowledgments}

This work was supported by funds provided by DGESIC under the
Research Projects No. HP1988--0040 and PB97--1218, and by the
Research Grant PRAXIS/2/2.1/FIS/286/94.

\section*{Appendix A}
\setcounter{equation}{0}
\renewcommand{\theequation}{A.\arabic{equation}}

In this appendix, we will discuss the freedom available in the
choice of coefficients in the series (\ref{Tas}). We will first
show that all the coefficients in these series can be determined
from the knowledge, at all instants of time $\tau$, of the
values of $v_0^c$, $V_0$, and $y_0^{(k)}$, as well as the value
of $Y_0$ when $\kappa=-1$. In order to demonstrate this
statement, let us call $\Gamma_n$ the set formed by $v_0$,
$y_0^{(\kappa)}$ and all the coefficients $\{V_m,Y_m,P_m,Q_m\}$
with $m\leq n$, where $n$ is an integer. By expanding the
dynamical equations (\ref{vdot})-(\ref{Pydot}) in powers of $r$,
one can check that the coefficients with subindex equal to $n+1$
can be determined from the knowledge of $\Gamma_n$ at all values
of $\tau$. Consequently, all the information needed to fix the
series is contained in $\Gamma_1$. Let us now analyse the
freedom available in the choice of this set of coefficients. One
can check that the lowest-order contributions to
$\partial_{\tau}{v}$ and $\partial_{\tau}{y}$ in equation
(\ref{vdot}) and (\ref{ydot}), respectively, determine $P_0$ and
$Q_0$ in terms of the constant $y_0^c$ and the derivatives of
$V_0$ and $Y_0$ with respect to $\tau$ when $\kappa=-1$, or as
functions of $y_0^{(1)}$ and the derivatives of this coefficient
and $V_0$ if $\kappa=1$. In this last case, in addition, the
contributions of order $r^3$ to $\partial_{\tau}{P}_y$ turn out
to fix $Y_0$ in terms of $V_0$ and $y_0^{(1)}$, once $Q_0$ has
been found. This concludes the proof of our assertion.

Let us consider now the collection of coefficients that appear
in the series (\ref{Tas}) at a certain initial time $\tau_0$,
rather than as functions of time, and discuss the freedom that
exists in the choice of such initial data. It is not difficult
to check that the coefficients of these power expansions satisfy
functional relations which do not involve time derivatives, so
that they are not all independent at $\tau_0$. In particular,
for $\kappa=1$, the lack of contributions of order $r$ and $r^3$
in $\partial_{\tau}{P}_v$ implies that $V_1$ and $V_2$ are
proportional to $V_0$. In addition, the requirement that the
subdominant correction to $\partial_{\tau}{y}_0^{(1)}$ should be
of order $r^6$ turns out to fix the coefficients $Q_1$ and $Q_2$
as linear homogeneous functions of $Q_0$. Similarly, for
$\kappa=-1$, the vanishing of the derivative of $P_y$ with
respect to $\tau$ at orders $r$ and $r^3$ leads to a functional
dependence of the coefficients $Y_1$ and $Y_2$ on $y_0^c$, $Y_0$
and $V_0$. Taking into account this dependence and demanding
that $\partial_{\tau}{P}_v$ does not include contributions of
order $r$ and $r^3$, one can also determine the coefficients
$V_1$ and $V_2$ as functions of $Y_0$ and $V_0$. In general,
more complicated relations appear when higher-order coefficients
are considered. As a consequence of these relations, one can
prove by induction that, in order to determine the series
(\ref{Tas}) at $\tau_0$, it suffices to know at that moment the
values of $v_0^c$, $y_{0}^{(\kappa)}$ and all the sets of
coefficients $\{V_{3n},Q_{3n},P_{3n},Y_{3n}\}$ with $n\geq 0$.

\section*{Appendix B}
\renewcommand{\theequation}{B.\arabic{equation}}

In this appendix, we will prove that the reduced Hamiltonian
$H_R$ is differentiable on phase space once one adopts the
boundary conditions (\ref{A2S}) [or (\ref{ANS}) if the angular
momentum vanishes], (\ref{conP}), and (\ref{conI}) [where
$v_{\infty}=v_{\infty}^c$]. The variation of the reduced
Hamiltonian is given by
\begin{equation}\label{varH}
\delta H_R=\frac{e^{-w_{\infty}}}{4}\int_0^{\infty}dr\delta\bar{H}
\left(2+c_{\theta}^2\bar{E}_{\infty}F^2_{\infty}\int_r^{\infty}
\frac{ds}{s^3}\bar{E}[s]\right),\end{equation}
where $\bar{H}$ is the function on phase space defined in
equation (\ref{barH}). The integrals that appear in the term in
parentheses are all convergent for $r>0$, owing to our boundary
conditions. In addition, from a variation of the canonical
momenta $P_v$ and $P_y$, one obtains
\begin{equation}
\delta\bar{H}=\frac{4}{r}(P_vr^2e^{2y}\delta P_v+P_y\delta P_y).
\end{equation}
Therefore, the Hamiltonian is differentiable with respect to
these momenta if and only if the integral over $r$ in the
expression of $\delta H_R$ converges (both at $r=0$ and at
infinity) for all possible variations of $P_v$ and $P_y$. Given
the asymptotic behaviour (\ref{conP}) and the conditions on the
axis, equation (\ref{ANS}) or equation (\ref{A2S}), the
admissible variations of $P_v$ and $P_y$ turn out, in general,
to be of the same order as the momenta themselves. It is then a
simply exercise to check that the studied integration over $r$
leads, in fact, to a well defined variation of the reduced
Hamiltonian.

The analysis of the variations of $v$ and $y$ is more
complicated. In this case, one gets
\begin{equation}
\delta\bar{H}=\frac{1}{r}\left[4P_v^2r^2e^{2y}\delta y-
(\partial_rv)^2e^{-2y}\delta y+r^2\partial_ry\partial_r (\delta
y)+\partial_rve^{-2y}\partial_r(\delta v)\right].
\end{equation} Note that this variation contains the
radial derivatives of $\delta v$ and $\delta y$. To get rid of
these derivatives, one must perform an integration by parts.
Consequently, the variation of the reduced Hamiltonian splits
into a surface term and an integral expression. The surface term
is
\begin{equation}\label{surf}
\frac{e^{-w_{\infty}}}{4r}(r^2\partial_ry\delta y+\partial_rv
e^{-2y}\delta v)\left.\left(2+c_{\theta}^2\bar{E}_{\infty}F^2_
{\infty}\int_r^{\infty}\frac{ds}{s^3}\bar{E}[s]\right)
\right|_0^{\infty},\end{equation}
where we have employed the notation $f|^b_a=f(b)-f(a)$. The
integral contribution to $\delta H_R$, on the other hand, can be
obtained by replacing $\delta\bar{H}$ on the right-hand side of
equation (\ref{varH}) with
\begin{equation}
4P_v^2re^{2y}\delta y-(\partial_r v)^2\frac{e^{-2y}}{r}\delta
y-\partial_r(r\partial_r y)\delta y-\partial_r\left(\partial_r
v\frac{e^{-2y}}{r}\right)\delta v,\end{equation} and adding to
the result the factor
\begin{equation}\label{int}
\frac{e^{-w_{\infty}}}{4}\int_0^{\infty}dr\frac{c_{\theta}^2}
{r^4}\bar{E}_{\infty}F^2_{\infty}\bar{E}[r]\left[r^2\partial_r
y\delta y+\partial_rv e^{-2y}\delta v\right].\end{equation}

The differentiability of the reduced Hamiltonian then requires
that the surface term (\ref{surf}) vanishes and that the
integrals over $r$ that determine the variation of $H_R$ are
convergent for all possible values of $\delta v$ and $\delta y$.
Actually, according to the boundary conditions (\ref{conI})
(with $v_{\infty}=v_{\infty}^c$), one has that $\delta
v=\bar{o}(1)$ and $\delta y=o(1)$ when $r\rightarrow\infty$.
Here, we have imposed that the variations of $v$ preserve the
value of $v_{\infty}^c$, since this is a given constant.
Similarly, from equation (\ref{ANS}) one concludes that, when
$c_{\theta}$ vanishes, the acceptable variations of our fields
display the behaviour $\delta v=O(r^2)$ and $\delta y=O(1)$
around the axis. Finally, in the presence of spin, one obtains
from equation (\ref{A2S}) that $\delta v=O(r^4)$ at $r=0$,
whereas the variation of $y$ admits two different types of
behaviour, depending on whether $y_0^{(\kappa)}$ is a fixed
constant or not. In the first case ($\kappa=-1$), one arrives at
$\delta y=O(r^2)$; in the second case ($\kappa=1$), one gets
$\delta y=O(1)$, because now it is possible to vary the limit of
$y$ on the axis. With this information about the variations of
the fields and our boundary conditions, it is not difficult to
check that the reduced Hamiltonian is indeed differentiable with
respect to $v$ and $y$. We therefore conclude that our boundary
conditions guarantee that $H_R$ is differentiable on phase
space, as we wanted to show.


\begin{thebibliography}{99}
\bibitem{di} This notation differs from that used in \cite{me}
\bibitem{ls} See, e.g., \cite{sou,dislo}, where the
Einstein-Cartan formalism of gravity was employed
\end{thebibliography}
\end{document}